\documentclass[12pt]{iopart}

\usepackage{enumerate}
\usepackage{enumitem}
\usepackage{graphicx}
\usepackage{dcolumn}
\usepackage{bm}
\usepackage{graphicx}
\usepackage{amssymb}
\usepackage{epstopdf}
\usepackage{color}
\usepackage[usenames,dvipsnames,svgnames,table]{xcolor}
\usepackage{placeins}
\usepackage{array}

\usepackage[english]{babel}
\usepackage{xspace}

\newcommand{\YBCO}{$\mathrm{YBa_{2}Cu_{3}O_{7}\;}$\xspace}

\definecolor{orange}{rgb}{1,0.5,0}

\FloatBarrier

\begin{document}

\title{High-Tc Superconducting Quantum Interference Filters (SQIFs) made by ion irradiation}

\author{S. Ouanani$^1$, J. Kermorvant$^2$, C. Ulysse$^3$, M. Malnou$^4$, Y. Lema\^itre$^1$, B. Marcilhac$^1$, C. Feuillet-Palma$^4$, N. Bergeal$^4$, D. Cr\'et\'e$^1$, J. Lesueur$^4$}

\author{$^1$Unit\'e Mixte de Physique CNRS-Thales, 1 Av. A. Fresnel, 91767 Palaiseau, France}
\author{$^2$Thales Communications and Security, 4 Av. des Louvresses, 92622 Gennevilliers, France}
\author{$^3$Laboratoire de Photonique et de Nanostructures LPN-CNRS, Route de Nozay, 91460 Marcoussis, France}
\author{$^4$Laboratoire de Physique et d\'{\'E}tude des Mat\'eriaux-CNRS-ESPCI ParisTech-UPMC,
PSL Research University, 10 Rue Vauquelin - 75005 Paris, France}

\date{\today}

\ead{jerome.lesueur@espci.fr}

\begin{abstract}

Superconducting Quantum Interference Filters (SQIFs) are arrays of superconducting loops of different sizes including Josephson Junctions (JJ). For a random distribution of sizes, they present a non-periodic response to an applied magnetic field, with a large transfer function and a magnetic field sensitivity potentially improved with respect to that of a single SQUID. Such properties make SQIFs interesting devices to detect the magnetic component of electromagnetic waves at microwave frequencies. We have used the highly scalable technique of ion irradiation to make SQUIDs and SQIFs based on commercial \YBCO films, and studied their properties. Both display optimum performances as a function of temperature and bias current, that can be understood in the frame of numerical simulations that we developed. The role of asymmetries and dispersion in JJ characteristics (routinely found in High Tc Superconductors technologies) is also studied. We have found that both do not impede the existence of a SQIF effect but play a role on the emergence of the optimal point. We finally present results on SQIF made with 2000 SQUID in series, showing a transfer function $dV/dB\sim1000 V/T$.

\end{abstract}

\section{Introduction}
Since the pioneering works of Carelli et al\cite{Carelli} and Oppenl{\"a}nder et al\cite{Oppenlander:2000hz,Oppenlander:2003fc}, a lot of work has been devoted to the development of Superconducting Quantum Interference Filters (SQIFs) for sensitive absolute magnetometry or for RF applications. Indeed, since the response of the SQIF is non-periodic in magnetic field, there is no need of a feed-back loop to maintain a fixed functioning point. Not only the absolute value of the magnetic field can be therefore measured, but the band-width of the device is not limited by the feed-back electronics anymore (typically a few MHz usually \cite{Clarke:2005tz}). In addition, one can chose the arrangement of the SQUIDs in the array to match the impedance of the device to the read-out system.

This paves the way for RF applications of SQIFs, such as compact low noise amplifiers and sub-wavelength broad band antennas (for a review see Mukhanov et al\cite{Mukhanov:2014ig}). 
In the recent years, different architectures and geometries of SQIFs have been explored, 1D in series and parallel configuration \cite{Oppenlander:2003fc}, 2D in series/parallel configurations \cite{Oppenlander:2003iq}, using conventional SQUIDs and bi-SQUIDs\cite{Kornev:2009bq} to obtain the best performances in terms of sensitivity, linearity and noise. For high-frequency applications, very interesting results have been obtained such as 8-15 GHz antennas in the near field\cite{Prokopenko:2015dx}, and RF amplifiers in the 12 GHz range\cite{Mukhanov:2014ig}. These performances have been obtained using low $T_{c}$ materials, mainly Nb, operating at liquid-He temperature. A lot of applications cannot afford the related costly and power consuming cryogenics which is needed in that case. High $T_{c}$  Superconductors (HTSc) such as \YBCO (YBCO) appear therefore as promising candidates to make SQIFs operating at high temperature, easily accessible with compact and low energy consumption cryocoolers. 
HTSc SQIFs have been fabricated for DC magnetometry applications \cite{Schultze:2003gl,Schultze:2003fb,Caputo:2005cq} and RF ones\cite{Caputo:2006jg,Caputo:2007cs,Snigirev:2007ih}, using mainly bicrystal grain-boundary JJ. By construction they all lie on the same line, which put stringent constraint on the circuit design. If 1D arrays can be easily realized, it is more challenging to achieve a high performance 2D HTSc SQIF along a geometrically one-dimensional grain boundary \cite{Schultze-2006,Mitchell-2016}.

 An alternative technology to make large arrays of HTSc SQUIDs is the irradiation technique developed by the San Diego group\cite{Chen:2004hba,Katz:1998gh}, and more recently by Bergeal et al\cite{Bergeal:2005jna,Bergeal:2007jc}. Starting from a commercial YBCO film, the circuit and the JJ are patterned by ion-irradiation through photoresist masks. The induced defects within the HTSc material lower the $T_{c}$ locally to fabricate SNS (Superconductor-Normal metal-Superconductor) JJ , and make it insulating at high fluence to draw the circuit. Superconducting devices such as THz Josephson mixers were recently realized by this method\cite{Malnou:2014cp,Malnou:2012gt}. Large 2D arrays of JJ\cite{Cybart:2009fca} and SQIFs\cite{Cybart:2008ff,Ouanani:2014cu} have been developed successfully. 
 
 In this article, we present the properties of  SQIFs made by ion irradiation, and explain the emergence of an optimal operating point (bias current and temperature). A deep understanding of the operation of SQIF devices is necessary for future performance improvement. For that, the first step is to figure out how a single SQUID made by ion irradiation works, since it is the key element of SQIFs. We performed numerical simulations to describe the characteristics of single SQUIDs that we measured experimentally. We then extended them to a small series SQIF, and drew conclusions about its optimal working point, and its sensitivity to dispersion in junction characteristics. Finally, a 2000 SQUID in series SQIF is presented with a transfer function in the $1000 V/T$ range.

\section{HTSc Josephson circuits made by ion irradiation}

\begin{figure}[h]
\includegraphics[width=\textwidth]{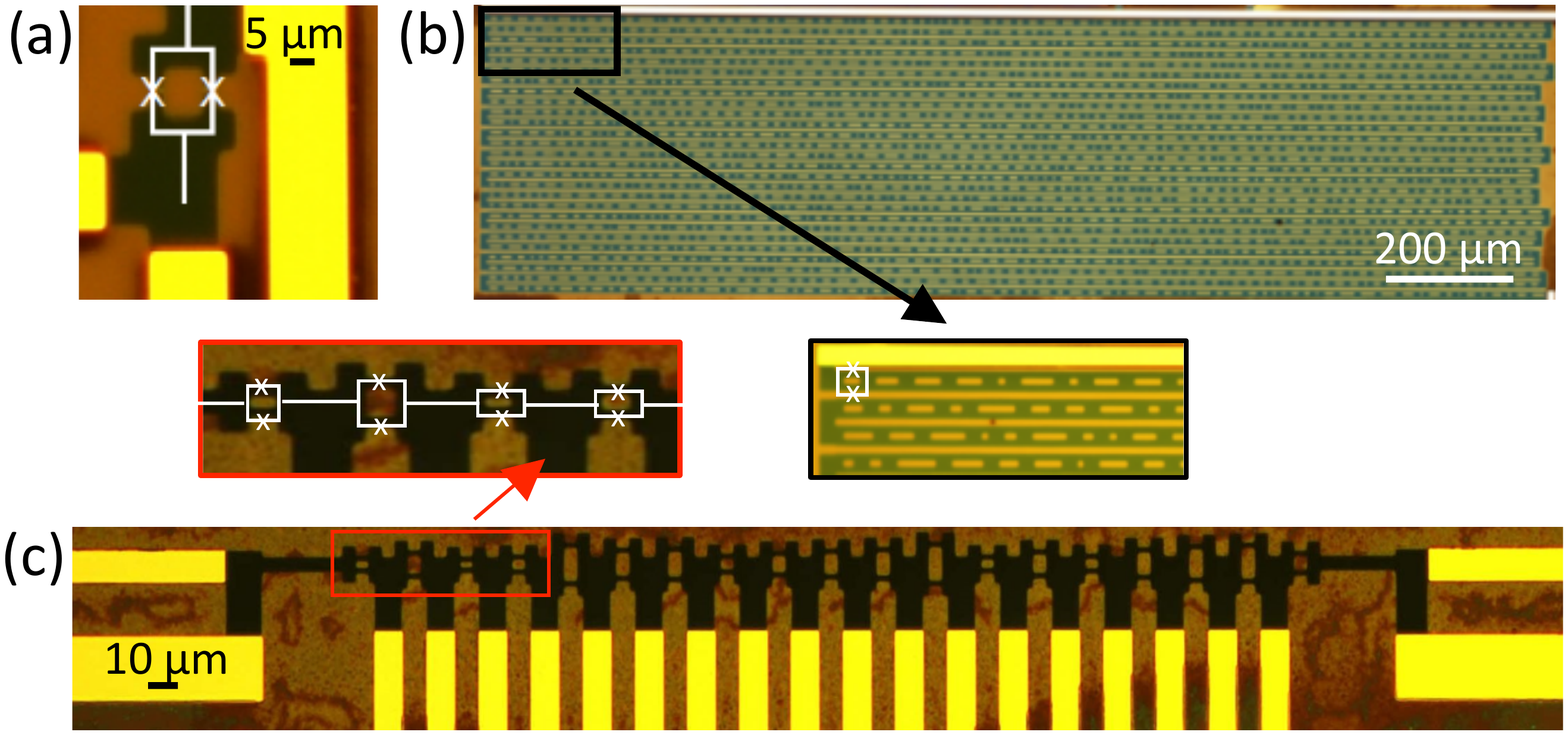}
\caption{Optical pictures of devices made by ion irradiation. White lines and crosses superimposed on pictures sketch the SQUID loops and the Josephson Junctions respectively. (a) $30\ {\mu}m^{2}$ SQUID with $2\ {\mu}m$ wide branches corresponding to geometric inductance $L_{geo}$=11 pH. (b) SQIF with 2000 SQUIDs in series, with $2\ {\mu}m$ wide branches and areas ranging from $6$ to $60\ {\mu}m^{2}$ ($L_{geo}$=4.9 to 22 pH). (c) 18 SQUIDs in series with $2\ {\mu}m$ wide branches and areas ranging from $5.6$ to $40\ {\mu}m^{2}$ ($L_{geo}$=5.1 to 13.3 pH). The first two devices (a) and (b) were realized on a same chip and the (c) on another one.}
\label{Figure1}
\end{figure}

In this study, we used commercial 150 nm thick YBCO films in-situ covered by a 100 nm gold layer\cite{Ceraco}. To start with, the latter is removed from the surface of the sample with the exception of the contact pads. The superconducting circuit is created by a first $110\ keV$ oxygen ions irradiation at a fluence of $5\times10^{15}\:ions/cm^{2}$ performed through a patterned photoresist. $40\ nm$ wide slits located across  superconducting microbridges are then patterned in a PMMA resist by e-beam lithography. A second oxygen ions irradiation at lower fluence (typically $3\times10^{13}\:ions/cm^{2}$) is performed to locally decrease $T_{c}$ and create SNS JJs. Figure~\ref{Figure1} shows optical pictures of different devices such as a single SQUID (a), a 2000 loops series SQIF (b) and an array of 18 SQUIDs in series that can be measured individually (c). The first two devices were achieved on one chip and the last one, on another substrate. Details of the fabrication process can be found elsewhere \cite{Bergeal:2005jna,Bergeal:2007jc,Malnou:2014cp,Malnou:2012gt,Ouanani:2015tw}. It is worth noting that same irradiation conditions were kept for all devices. Moreover design does not include large flux focussing areas around the loops or any transformers.

Under such conditions, Josephson coupling occurs at around $T_{J}\sim70 K$, and extends over typically $15 K$ (see Figure~\ref{Figure2} (a) for a single SQUID). Below, a flux flow regime takes place, for temperatures smaller than the critical temperature  of the irradiated part called $T_{c}^{'}$. 

All  measurements presented here were done in a magnetically unshielded pulse-tube cryocooler. The magnetic field was applied perpendicular to the sample by an external Helmoltz coil system. We choose to not use a magnetic shield in order to explore the effect of a real environment on devices.

\section{Characteristics of a DC-SQUID made by ion irradiation}

\begin{figure}[h]
\includegraphics[width=11cm]{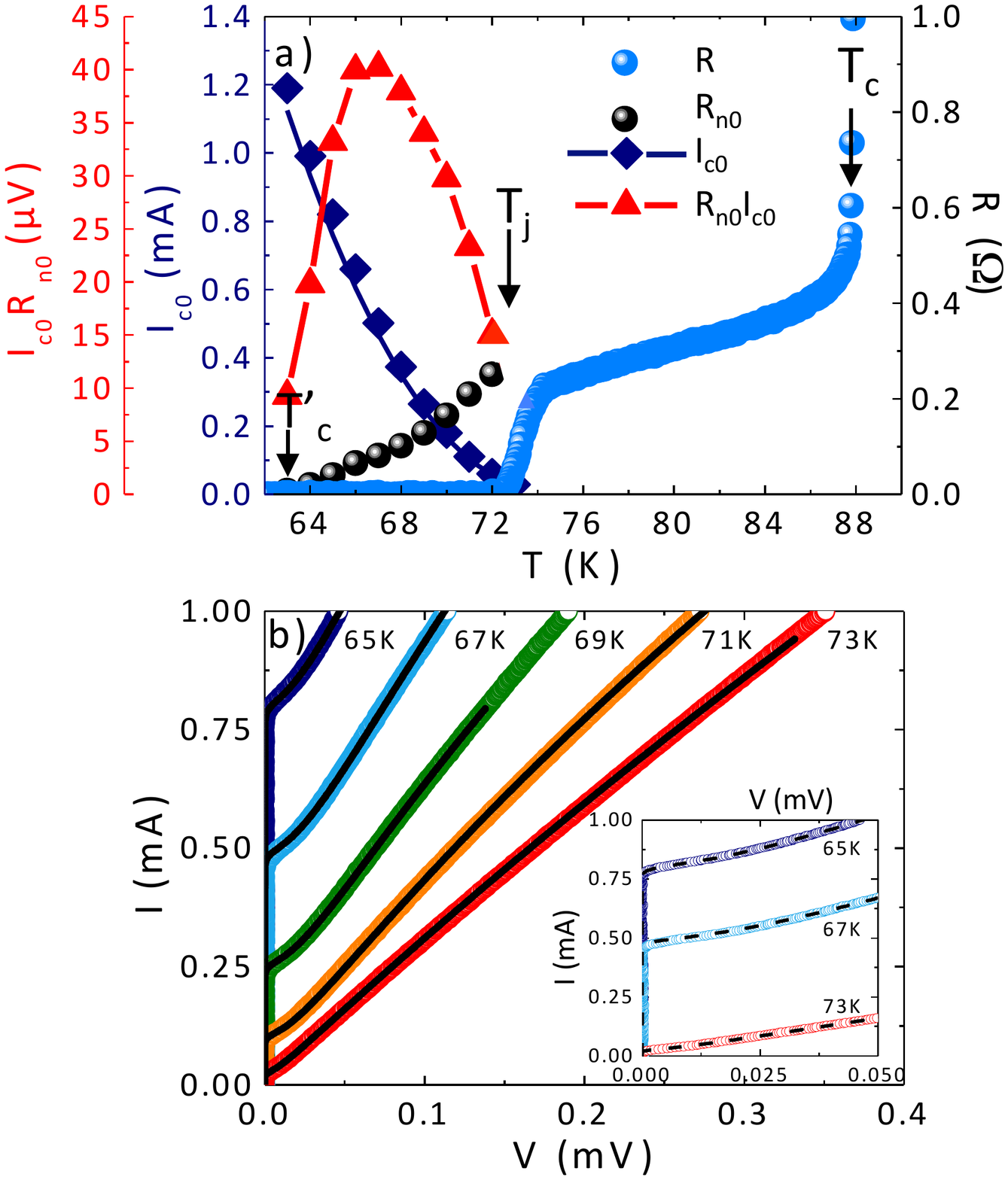}
\caption{DC characteristics of the SQUID. (a) Resistance (right scale) and critical current (left scale) curves vs. temperature showing the different characteristic temperatures $T_{c}$, $T_{J}$ and $T_{c}^{'}$ (see text) and the $I_{c0}\:R_{n0}$ product (extreme left scale) showing a pronounced dome shape. The resistance $R_{n0}$ (see main text) is reported for $T<T_{J}$. (b) I-V characteristics in the Josephson regime, for different temperatures. Color symbols are experimental data and black broken lines are fits with the modified RSJ model (see text). Inset shows a zoom around the critical current area for T=65, 67 and 73K.
}
\label{Figure2}
\end{figure}

First, we  measured  the  resistance of the device as a function of temperature as depicted in Figure \ref{Figure2}.a. The  device  exhibited  two  distinct  superconducting  transitions at 88K and 73K. The highest transition refers to that of the electrodes (the same as the unprocessed film) and the second to $T_{j}$, below which the Josephson regime starts. We then measured the I-V characteristics for the SQUID, shown in Figure \ref{Figure2}.b in zero magnetic field (B=0) and for several temperatures. In the Josephson regime, I-V characteristics of the SQUID follow a modified Resistively Shunted Junction (RSJ) model, with a non-linear normal state resistance $R_{n}=R_{n0}+\chi\:I$\cite{Malnou:2014cp,Katz:2000gf}, where $I$ is the bias current, $R_{n0}$ the resistance in the regular RSJ model \cite{McCumber1,Stewart}, and $\chi$ is a parameter determined by fitting the experimental data.
Given the rather high operation temperature $T$, thermal smearing has to be taken into account in the fit through the parameter $\Gamma=4\pi\:k_{B}T/I_{c0}\Phi_{0}$  ($I_{c0}$ is the SQUID critical current for B=0,  $\Phi_{0}$ the flux quantum and $k_{B}$ the Boltzmann constant). Fits of the I-V characteristics (see Figure~\ref{Figure2} (b)) give access to $R_{n0}$ and $I_{c0}$, and therefore to the $I_{c0}\:R_{n0}$ product which has a dome shape as a function of temperature\cite{Malnou:2014cp,Ouanani:2015tw} (see Figure~\ref{Figure2} (a)). This is an essential feature of HTSc JJ made by ion irradiation.  This shape is due to specific dependences of the critical current and the normal resistance versus temperature. Indeed, $R_{n0}$ decreases linearly with temperature to reach a null value at $T_{c}^{'}$ (the intrinsic critical temperature of the irradiated part) and $I_{c0}$ increases quadratically with temperature : this leads to  a maxima of the  $I_{c0}\:R_{n0}$ product.

\begin{figure}[h]
\includegraphics[width=12cm]{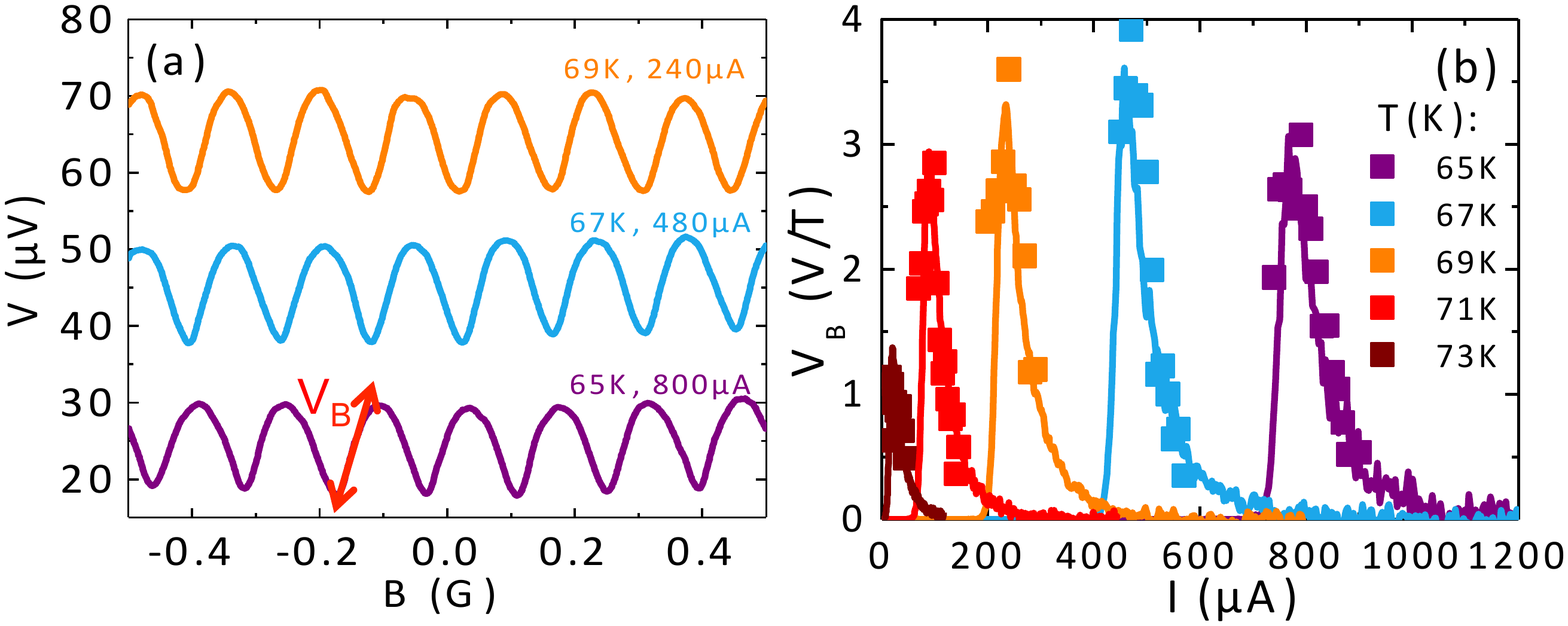}
\caption{(a) SQUID voltage modulations as a function of magnetic field B for different temperatures at the optimum current biasing in an unshielded environment. (b) Transfer function $V_{B}$ as a function of the bias current for different temperatures. Symbols are experimental data and solid lines are the result of our simulations based on the RSJ model (see text).  
}
\label{Figure3}
\end{figure}

Figure~\ref{Figure3} (a) shows typical voltage modulations of the SQUID as a function of an applied magnetic field $B$ measured in an unshielded environment for different temperatures. From these V-B curves, we extracted the evolution of the transfer function $V_{B}=\mid\partial\ V/\partial B\mid_{max}$ as a function of bias current (Figure~\ref{Figure3} (b)). As expected, all the curves show  a peak  in $V_{B}$ at a bias current approximately equal to  the  noise-free  critical  current  determined  by  the screening parameter $\beta_{L}=LI_{c0}/\Phi_{0}$ (where $L$ is the loop inductance)  for $\Phi=0.25\Phi_{0}$ \cite{TESCHE:1977ka}. This comes from the existence of thermal fluctuations which allow the SQUID to display a voltage modulation in magnetic field for bias currents lower than its critical current. 
$V$-$B$ curves recorded for several temperatures allowed us to extract the evolution of $V_{B}$ as a function of  temperature. A particularity is the presence of a maximum in temperature for  $V_{B}$ at $T_{max}\sim67 K$. To get more insight on this behavior, we compared our data with semi-empirical formulae given by Koelle et al\cite{Koelle:1999dy} and Enpuku et al\cite{Enpuku:1993kf,Enpuku:1993wy}, and with numerical simulations of the RSJ model. 

We performed numerical simulations starting from the fit of the I-V curves with the modified RSJ model (see above) to extract $R_{n0}$ and $I_{c0}$ for different temperatures. We used the Inductex software to calculate the inductance of the SQUID which includes geometric ($L_{geo}$) and kinetic inductance ($L_{K}$) since for these temperature this latter is not negligible anymore. $L_{K}$ depends on the superconducting penetration depth $\lambda_{L}$, which varies with temperature according to the Gorter-Casimir formula \cite{Tinkham:1996uv,Terai,Wolf} : $\lambda_{L}(T)=\lambda_{L}(0)/\sqrt{1-(T/T_{c})^\alpha}$ (with $\lambda_{L}(0)=220 nm$, $T_{c}=88 K$, $\alpha=2.1$). All parameters are listed in Table \ref{Table1}.
The resulting $V_{B}$ is shown in  Figure \ref{Figure3} (b), in very good agreement with experimental data as well as a function of the bias current than as a function of temperature. 

\begin{table}[h]
   \centering
   \begin{tabular}{|l|l|l|l|l|l|}
\hline
T (K) & $I_{c0} (\mu A)$ & $R_{n0} (\Omega)$ & $\chi (\Omega/A)$ & L (pH) & $\beta_{L}$ \\ 
\hline
65 & 820 & 0.0405 & 41.2 & 18 & 7.4\\
\hline
66 & 660 & 0.0605 & 42.2 & 18 & 5.9\\
\hline
67 & 502 & 0,0801 & 50 & 18.3 & 4.6\\
\hline
68 & 374 & 0.1015 & 60.2 & 18.8 & 3.5\\
\hline
69 & 265 & 0.1285 & 70.2 & 19.8 & 2.6\\
\hline
70 & 180 & 0.1655 & 75.2 & 20.8 & 1.9\\
\hline
71 & 111 & 0.2099 & 66.9 & 21.8 & 1.2\\
\hline
72 & 60 & 0.251 & 68.2 & 23.5 & 0.7\\
\hline
73 & 29 & 0.31 & 45.2 & 24.5 & 0.3\\
\hline
\end{tabular}
\caption{Parameters used for the simulations.}
\label{Table1}
\end{table}

\begin{figure}[h]
\includegraphics[width=12cm]{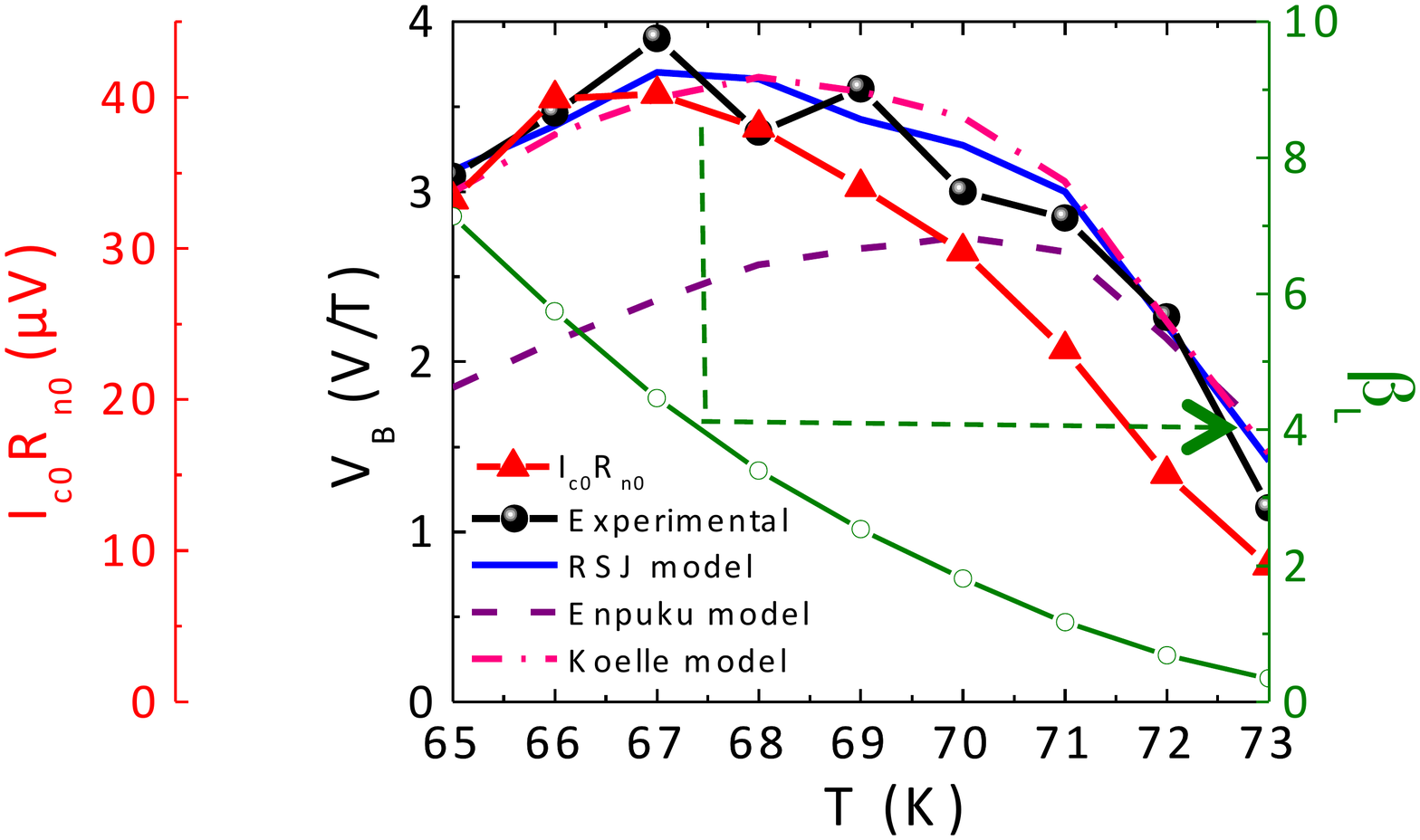}
\caption{$V_{B}$ of the SQUID at optimum bias current as a function of temperature (data black symbol, our simulation blue solid line, Enpuku model violet broken line, Koelle model pink dash-dotted line). Is also plotted $I_{c0}\:R_{n0}$ (red triangles, extreme left scale), and  $\beta_{L}$ (green open circles, right scale). }
\label{Figure4}
\end{figure}

The quite unusual behavior in temperature of $V_{B}$ is well reproduced in our simulations. In fact, this comes from the non-monotonic variation of the $I_{c0}\:R_{n0}$ product with temperature as mentioned above. In Figure~\ref{Figure4}, we plotted $V_{B}$  measured at optimum current (which is temperature dependent) as a function of temperature, and compared to our simulations and to the semi-empirical formulae. On the same graph are also reported the $I_{c0}\:R_{n0}$ product with its characteristics dome shape, and the value of $\beta_{L}$. The maximum of $V_{B}$ at 67-68 K comes from the convolution of $I_{c0}\:R_{n0}$ (maximum at 66-67 K) with the decreasing $\beta_{L}$. Indeed, $V_{B}$ is proportional to $I_{c0}\:R_{n0}/(1+\beta_{L})$ for small $\beta_{L}$\cite{Clarke:2005tz,TESCHE:1977ka,Ryhanen}, as seen in the Koelle's and Enpuku's formulae. Enpuku's expression describes correctly the experimental results in a restricted high temperature region, while the Koelle's one and our simulations do it on the full range. An important message here is that the optimum working point in current and temperature for the SQUID corresponds to $\beta_{L}\sim\ 4$ as marked by the dashed green arrow in Figure~\ref{Figure4}, and not for $\beta_{L}=1$. For all SQUIDs that we measured, the optimum as well as for $V_{B}$ than for $\Delta{V}=V_{max}-V_{min}$ is reached for high $\beta_{L}$ value. This seems to be inherent to SQUIDs made by ion irradiation. 


\section{Study of a small Series SQIF}

SQIFs are very promising devices for sensitive magnetometry, since their performance scale with the number of SQUIDs (linearly for the series configuration). However, the collective behavior is complex, since relevant parameters such as $\beta_{L}$ vary with the loop size on the one hand, and since deviations from the canonical fully symmetric SQUID and the dispersion in individual SQUIDs characteristics are unavoidable on the other hand. As a result, optimization of the array is not straightforward, and one would benefit from a thorough investigation of its detailed operation.
On the way of making SQIFs with thousands SQUIDs, we designed a 18 SQUIDs SQIF in a series configuration with loop sizes ranging from $5.6$ to $40\mu\ m^{2}$ ($L_{geo}$ from 5.1 to 13.3 pH), where the SQUIDs can be measured individually (see Figure~\ref{Figure1} (c)).  
For the whole device, we recorded the $V-B$ curves . 

\begin{figure}[h]
\includegraphics[width=12cm]{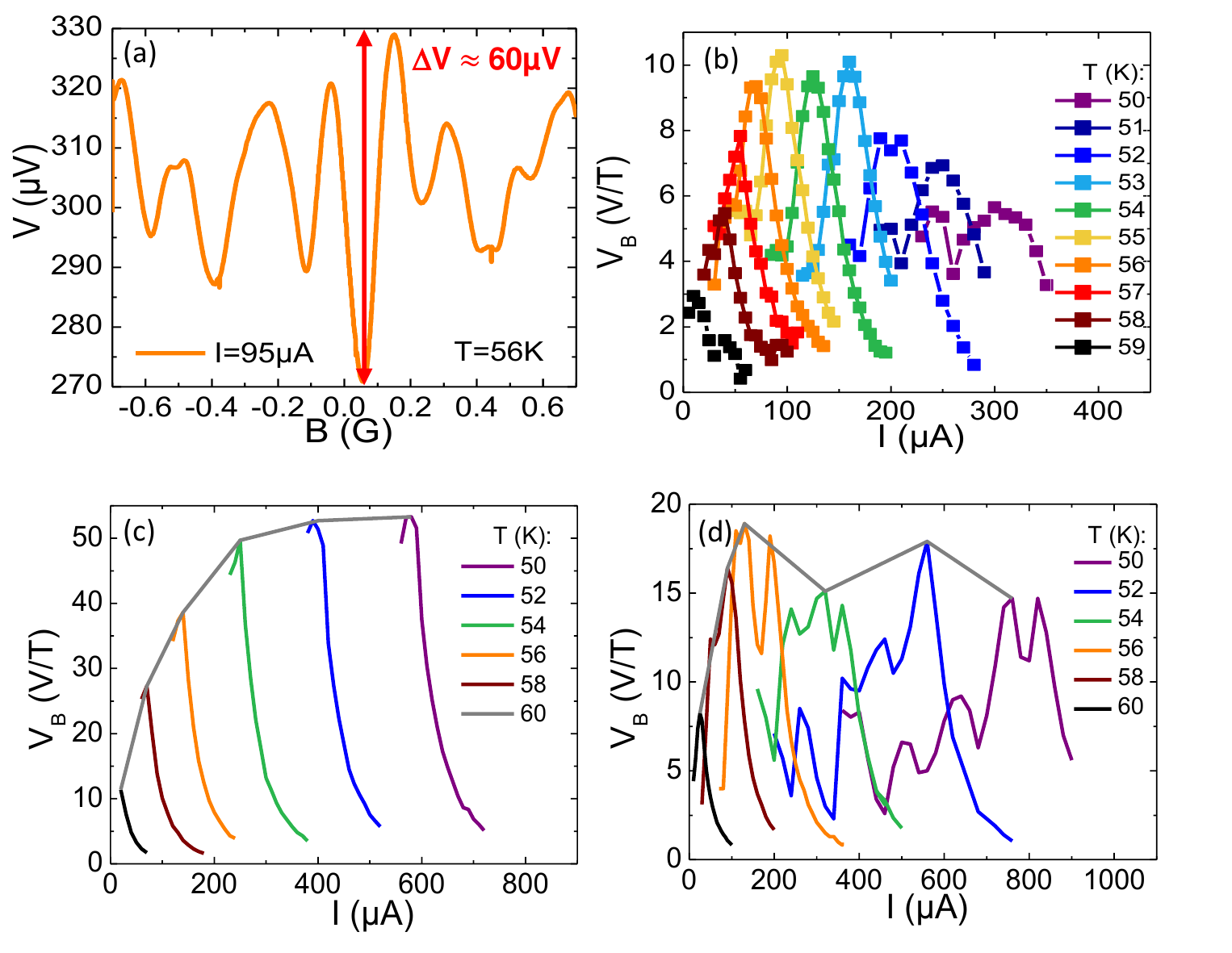}
\caption{(a) SQIF voltage modulation at 56K for the optimum bias current of $95\ {\mu}A$. (b) Measured transfer function $V_{B}$ as a function of bias current for different temperatures for the 18 SQUIDs SQIF. (c) Calculated transfer function as a function of bias current for different temperatures, without dispersion in the JJ characteristics (c) Idem with a dispersion in characteristics of $30\%$ (see main text). (d) } 
\label{Figure5a}
\end{figure}

As depicted in the Figure \ref{Figure5a} (a), a SQIF response is observed. It is weak, since the number of SQUIDs is moderate, but clearly seen, with an amplitude $\Delta\ V\sim 60 \mu\ V$ at the maximum. From V-B curves, we then extracted the parameter $V_{B}=\partial\ V/\partial\ B|_{max}$. Figure~\ref{Figure5a} (b) shows $V_{B}$ as a function of the bias current for different temperatures (similar results are obtained for $\Delta\ V$). The overall picture is similar to that of a single SQUID, with an optimum bias current for each temperature, and an optimum temperature at T=56K (this device displays a Josephson regime at temperatures lowers than the two others devices presented). To explain this behavior, we numerically modeled the V-B characteristics of a device with the same geometry as the experimental one.

We first simulated an array where all the JJ were identical. We used characteristics from a typical individual SQUID (namely $R_{n0}$, $I_{c0}$ and $\chi$ for different temperatures) and computed the SQIF response by summing the contribution of the SQUIDs with different loop sizes. We knowingly neglected mutual inductive coupling between SQUIDs, since the distance between nearest neighbors ($\sim14\mu\ m)$ is large enough. We also took into account the temperature dependence of both $\beta_{L}$ and $\Gamma$. 
The result is shown in Figure~\ref{Figure5a} (c) displaying $V_{B}$ as a function of the current bias for different temperatures. There is a clear discrepancy with experimental data, with a restricted range in current where the SQIF behavior can be observed, and an optimum temperature lower than the one experimentally obtained. We then introduced a dispersion in the JJ characteristics ($P=30\%$ in $I_{c0}$ and in $I_{c0}~R_{n0}$, where $P$ is equal to the standard deviation divided by the mean for a gaussian distribution). Recent statistical studies of ion-irradiated JJ in a Pb-shielded cryostat lead to a typical dispersion of $13\%$, while our measurements in unshielded environment show that dispersion varies from $7\%$ to roughly $30\%$ depending on the temperature\cite{Ouanani:2015tw}. We therefore used an upper bound estimate in our simulations.

By introducing a dispersion in JJ characteristics, we can notice (Figure~\ref{Figure5a} (d)) that the range in bias current where the SQIF voltage modulation is observed is larger at low temperatures but tends to be reduced at higher temperatures. Besides, a clear optimum occurs in temperature. It is worth noting that the optimal temperature is shifted toward higher temperature as compared to the case of null dispersion,and corresponds to that experimentally measured. Moreover, presence of dispersion tends to decrease the transfer function as compared to the ideal case. These results are consistent with computations and experiments reported in the literature\cite{Wu-2013,Berggren-2015}. To explain this behavior, it is necessary to come back to the behavior of the individual SQUIDs. 

\begin{figure}[h]
\includegraphics[width=12cm]{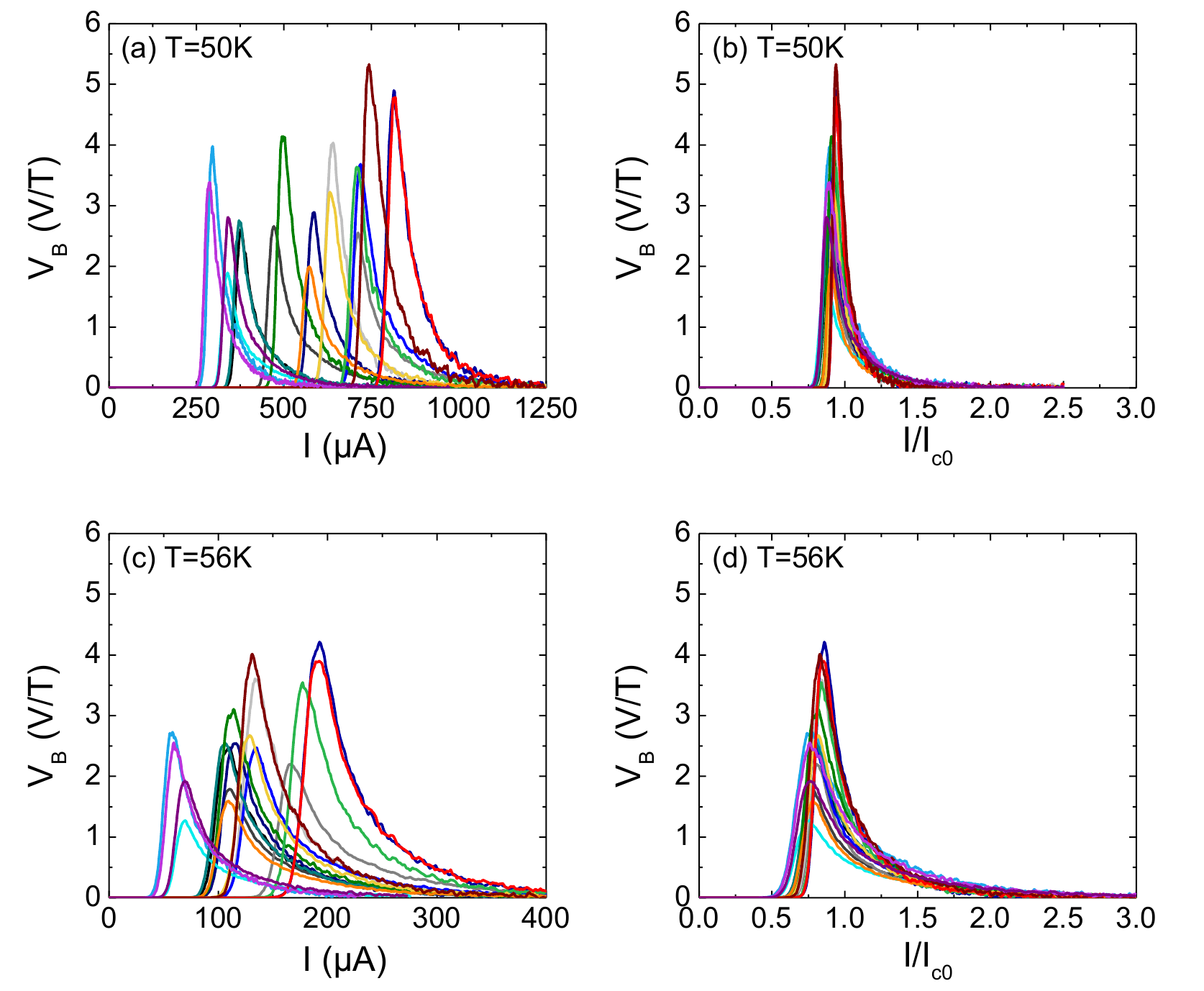}
\caption{Simulations of 18 SQUIDs with a dispersion in characteristics of $30\%$ (see main text). (Left panels) computed $V_{B}$ as a function of the bias current at two different temperatures : (a) 50K and (c) 56K. In the former case $1.9<\beta_{L}<6.7$ and $0.005<\Gamma<0.013$, and in the latter one $0.5<\beta_{L}<1.8$ and $0.02<\Gamma<0.062$. (Right panels) Same data normalized by $I_{c0}$. The peak at low temperature is narrower than the one at high temperature. 
} 
\label{Figure6}
\end{figure}

Figure~\ref{Figure6} shows the transfer functions vs bias current for 18 SQUIDs with different characteristics for two temperatures : 50K, panel (a) and 56 K, panel (c). Figures \ref{Figure6} (b) and (d) depict $V_{B}$ plotted as a function of the normalized current $I_{c}/I_{c0}$ for each temperature. 
These two temperatures correspond to two different SQUID working regimes. At low temperature (i.e. 50K), for a given current, only a very small fraction of SQUIDs works together which impedes the existence of a SQIF voltage modulation. By increasing the temperature, the overlap of voltage modulation of SQUIDs is more important raising the number of SQUIDs working together. Thus, a SQIF behavior can be observed with a well-defined voltage modulation. The reason is the following. At low temperatures, $\beta_{L}$ is large and the thermal parameter $\Gamma$ small for all SQUIDs. That means that each SQUID displays a voltage modulation in magnetic field in a restricted current range centered around its critical current (Figure \ref{Figure6} (b)). In the presence of dispersion in JJ characteristics, the different SQUIDs in the array will have distinct optimal currents and thus a collective behavior of the array is strongly disadvantaged. Moreover, even if the transfer function of individual SQUIDs is higher at low temperature, the potential SQIF response will be moderate, and will decrease with decreasing temperature, since fewer and fewer SQUIDs contribute. On the contrary, at high temperature, each SQUID displays a voltage modulation in magnetic field on an extended range of current centered around its critical current, which leads to a collective behavior of the array. For a given current, the situation with SQUIDs operating in low-beta regime favors the contribution of many SQUIDs to the SQIF response, but the amplitude of each SQUID voltage modulation being smaller at high temperature, $V_{B}$ decreases as the temperature increases. 
\begin{figure}[h]
\includegraphics[width=12cm]{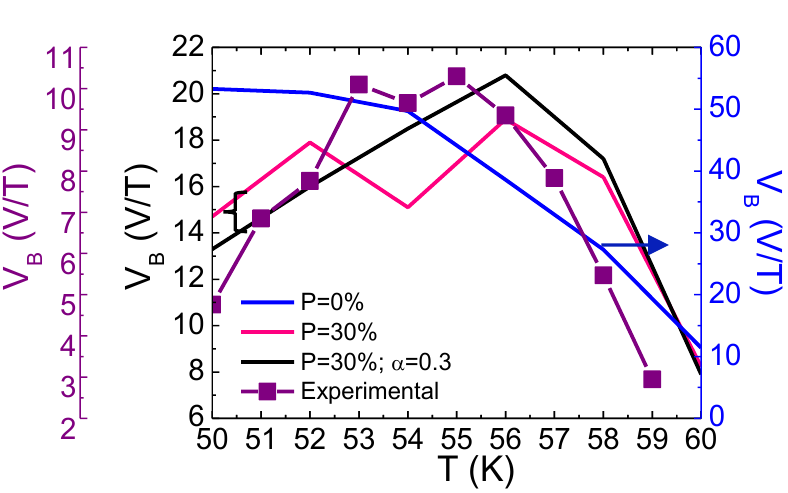}
\caption{Transfer function of the SQIF at the optimum bias current as a function of temperature. Black purple symbols are data (extreme left scale). Solid lines are computed results without dispersion (blue line and right scale), with $30\%$ dispersion in the JJ characteristics for symmetric SQUIDs (pink line), and with in addition $30\%$ asymmetry in the critical currents of the two JJ in each SQUIDs (black line).} 
\label{Figure5b}
\end{figure}
This is why an optimum in temperature is observed as summarized in Figure~\ref{Figure5b}, where both experimental data and simulations with, and without dispersion are shown. 

We can go one step further and include an asymmetry in the critical currents of the JJ in single SQUIDs. It is worthwhile reminding that an asymmetry in the critical current of the SQUID is equivalent to a non-zero magnetic field within the SQUID loop. We introduced a $30\%$ difference in the critical currents for all SQUIDs in our simulations, corresponding to the measured dispersion for single JJ\cite{Ouanani:2015tw}. The result is the curve shown in Figure~\ref{Figure5b} (black solid line), a bit closer to our experimental data. In the same spirit of the previous arguments, asymmetry will impair the collective behavior at low temperature when few SQUIDs contribute to the SQIF response. Moreover, the shift in V-B curves of SQUID has more dramatic effect in this situation. Individual SQUIDs do not respond simultaneously, which leads to the the reduction, or even the disappearance of the SQIF voltage modulation. This effect is less deleterious at high temperature. As a consequence, a more pronounced optimal temperature appears in the presence of asymmetry in critical current. Experimentally, since we perform our experiment in an unshielded environment to stay close to the real applications of SQIFs, we cannot distinguish between potential asymmetry in JJ characteristics and uncontrolled stray fields which would have the same effect on the SQIF response. 

As a summary, these results and their analysis based on our RSJ simulations show that SQIF effect in our devices is quite robust against dispersion in JJ characteristics and asymmetries. Nevertheless, dispersion acts on the operating point by shifting it toward high temperature and low bias current.

\section{Characteristics of 2000 SQUID in Series SQIF}

\begin{figure}[h]
\includegraphics[width=12cm]{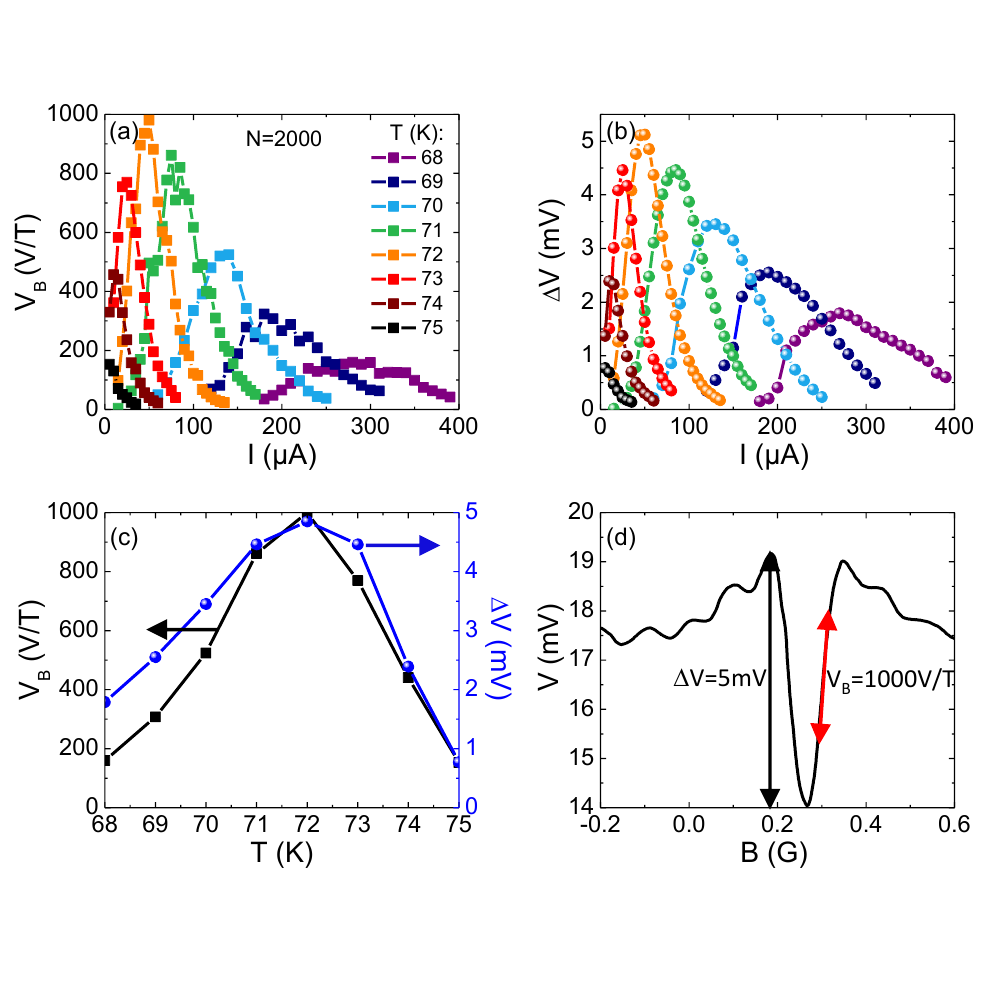}
\caption{Transfer function $V_{B}$ (a) and SQIF voltage modulation amplitude $\Delta V$ (b) as a function of bias current for different temperatures of the 2000 SQUIDs SQIFs in a series configuration. (c) $V_{B}$ (black symbols and left scale) and $\Delta V$ (blue symbols and right scale) at the optimum bias current for different temperatures. (d) Maximum SQIF response $V(mV)$ as a function of an applied magnetic field in an unshielded environment ($T=72K$, and $I=50 \mu A$).
} 
\label{Figure7}
\end{figure}

We made series arrays of 2000 SQUIDs with loop sizes ranging from $6$ to $60\ {\mu}m^{2}$ ($L_{geo}$=4.9 to 22 pH) and $2\ {\mu}m$  wide branches to make a SQIF (see Figure~\ref{Figure1} (b)), which operates between 65K and 75K. $V_{B}$ and $\Delta V$ as a function of bias current for different temperatures are reported in Figure~\ref{Figure7} (a) and (b) respectively, and display optimum operating conditions as (see Figure~\ref{Figure7} (c)). This is coherent with previous experiment and simulations reported above. The optimum temperature in both cases is $72K\pm1K$, where $V_{B}\sim1000 V/T$, and $\Delta V\sim5 mV$. These numbers compare favorably with previous reports in the literature. Using the ion-irradiation technique, Cybart \textit{et al} reported $V_{B}\sim100 V/T$ and $\Delta V\sim3 mV$ for 280 SQUIDs \cite{Cybart:2008ff}. Oppenl\"ander and co-workers using grain-boundary HTSc JJ published $V_{B}$ values between $3000 V/T$ and $8000 V/T$ and a $\Delta V$ in the $3 mV$ range for 100-200 SQUIDs in series\cite{Oppenlander:2003fc}. Mitchell \textit{et al} recently reported $1530 V/T$ in a 2D HTSc SQIF with 20 000 JJ\cite{Mitchell-2016}.

\begin{figure}[h]
\includegraphics[width=10cm]{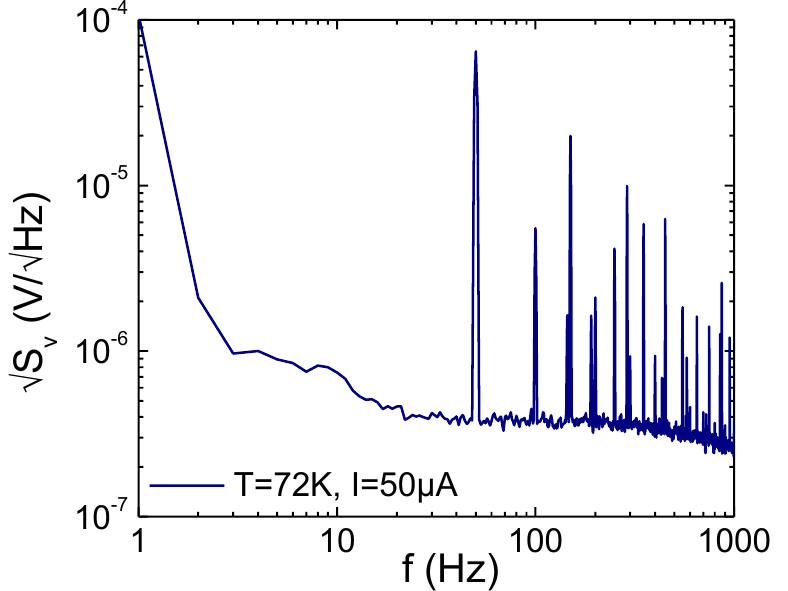}
\caption{Experimentally measured voltage output noise  spectrum $\sqrt{S_{V}(f)}$ vs frequency f of a 2000 SQUIDs in series-SQIF. The SQIF is operated at 72 K inside an unshielded cryostat. The  peaks  at $\approx$ 50 Hz  and  higher  harmonics are due to the grid.
} 
\label{Figure8}
\end{figure}

Low frequency noise measurements were performed on this device at the optimum point ($T=72 K$ and $I=50 \mu A$). Two low-noise amplifiers (noise level $\sim10 nV/\sqrt{Hz}$) were used in parallel, and only the correlated noise has been measured.
The obtained spectrum (Figure \ref{Figure8}) exhibits $1/f$ noise rising up at rather low frequency, namely below $\sim20 Hz$. Such a
value has been achieved with High $T_{c}$ devices only by using bias reversal techniques\cite{Oppenlander:2003fc}. Neither bias-reversal nor flux-locked loop electronics were used here.
The SQIF displays a white voltage noise level of $\sqrt(S_{V}( f ))\sim340 nV/\sqrt{Hz}$, in good agreement with theoretical predictions\cite{Oppenlander:2003fc,Likharev,Clarke1976}. Taking into account the obtained transfer function value  $V_{B}\sim1000 V/T$ for the bare SQIF (with no transformer), the voltage noise level can be translated into a magnetic field sensitivity of $\sim340 pT/\sqrt{Hz}$. Values found in the literature are generally lower\cite{Oppenlander:2003fc,Schultze:2003gl,Schultze:2003fb}.  However, the comparison is limited since the number of SQUIDs involved is more than an order of magnitude different, and the total impedance of the system is very different as well. Based on these results, an optimization of the SQIF design is being made to achieve better performances.

\section{Conclusion}

Both HTSc SQUIDs and SQIFs made by ion-irradiation present an optimum operating point in bias current and temperature. The specific dome like temperature dependence of the $I_{c0}R_{n0}$ product is important to understand the maximum in temperature. The SQIF behavior is more complex to analyze. Our simulations based on the RSJ model enable us to qualitatively reproduce our experimental results on small SQIF in a unshielded environment, provided we introduce a dispersion in the individual SQUIDs characteristics, coming from differences in JJ or from uncontrolled magnetic fields. Both spread and high temperatures favor a SQIF response from many SQUIDs, but with a small amplitude, while low temperature operation provides larger amplitudes but on a limited number of SQUIDs. This is the main reason for the optimum operating point observed. Finally, the 2000 SQUIDs in series SQIF displays interesting features in an unshielded environment, such as sensitivity in the $1000 V/T$ range and a low noise level. These performances look promising for the realization of performing HTSc SQIFs.

\section{Aknowledgments}

The authors thank Yann Legall for ion irradiations. This work has been supported by TRT through a CIFRE PhD fellowship, the T-SUN ANR ASTRID program, the Emergence program Contract from Ville de Paris and by the R\'egion Ile-de-France in the framework of the DIM Nano-K and Sesame programs. 

\clearpage

\end{document}